\documentstyle[12pt]{article}
\begin{document}
\title
{\LARGE \bf Fast Enumeration of Combinatorial Objects}
\date{}
\author{\large Boris Ryabko}
\maketitle

{\it Summary .} The problem of  ranking (or perfect
hashing) is well known
in Combinatorial Analysis, Computer Science, and Information Theory. There are
 widely used
 methods for ranking
  permutations  of numbers $\{1,2,...,n\},
n \geq1$, for ranking
  binary words of  length  $n$ with a fixed number
of ones and for many other combinatorial problems. Many of these
methods  have non-exponential memory size and the time of
enumeration $c_1 n^{c_2}$ bit operations per letter, where $c_1 >
0, \,c_2 \geq 1,\, n \to \infty$. In this paper we suggest a
method which also uses non-exponential memory size and has the
time of enumeration $O( (\log n)^{const})$ bit operations per
letter, $const > 0$, $n \to \infty$.

\vspace{1cm}

\section{Introduction}

The problem of ranking can be described as follows. We have a set
of combinatorial objects $S$, such as, say, the k-subsets of n
things, and we can imagine that they have been arranged in some
list, say lexicographically, and we  want to have  a fast method
for obtaining the rank of a given object in the list. This
problem is widely known in Combinatorial Analysis, Computer
Science and Information Theory (see [1,2]). Ranking is closely
connected with the hashing problem, especially with perfect
hashing and with generating of random combinatorial objects. In
Information Theory  the ranking problem is closely connected with
 so-called enumerative encoding [3], which  may be described as
follows: there is a set of words $S$ and an enumerative code has to
one-to-one encode every $s \in S$  by a binary word $code(s)$. The
length of the $code(s)$ must be the same for all $s \in S$.
Clearly, $|code (s)|\geq \log |S|$. (Here and below $\log
x=\log_{2}x)$.)

The simplest method of coding is to store all words from S and all
words $code (s) , s \in S$, in the memory of the encoder and the
decoder. In this case the time for encoding and decoding is
proportional to $\log |S|$ and, obviously has a minimal value
within a multiplicative constant when $|S|$  grows. However, the
memory size of the encoder and decoder increases exponentially (as
a function of the word length) just because they need to store all
words $s \in S$ and $code (s), s \in S$. Fortunately, for many
interesting and important problems of enumeration there exist
methods which do not use exponential memory size, see, for
example, [1,2]. We consider two examples of such problems:
enumeration of permutations and enumeration of the set of  binary
words with a given number of ones. These examples are well  known
in Combinatorial Analysis. Note that the second problem is very
important for Information Theory where it forms the basis for many
data compression codes. The first  code, which does not use
exponential memory size, was developed by Lynch [4], Davisson [5]
and Babkin [6]  (see also [2]).
For this code the time of encoding and decoding per letter is more than 
$const \cdot n$ bit operations. This also holds  for the time of
encoding end decoding for known methods for ranking of permutations.

In this paper we suggest a new method for ranking (or enumerative
encoding) for which the time  of encoding and decoding is
$O(\log^{const}n)$ bit operations per letter. This method is based
on the divide-and-conquer principle and uses the
Sch\"{o}nhage-Strassen method of fast multiplication. As mentioned
above, the proposed method is better than the known ones when
there exists an algorithm with non-exponential memory size. The
suggested method allows the exponential growth of the speed of
encoding and decoding for all combinatorial problems of
enumeration which are considered, for example, in [1] and [2]
including the enumeration of permutations, compositions and
others.

The next part describes  the main idea of the proposed method. The
descriptions of encoding and decoding are given in the parts 3 and
4, respectively.

\section{The Main Idea}

The simplest but important example of the problem of ranking (and
enumerative encoding) is the problem of integer translation from
one radix to another.
 We will use  this
example to represent the main idea of the proposed method.

Consider the task of  translation of an integer from a radix $m (m \geq 2)$ to the
binary system. Let there be given an integer $x_{1}x_{2}...x_{n}$, $n \geq 1$,
in the number system $m$. A ``common'' method of translation is based on the
following equality:
$$ code (x_{1}...x_{n})= \sum_{i=1}^{n} x_{i} m^{n-i}$$

Instead of this formula we can use the well- known Horner's scheme :
\begin{equation}
code (x_{1}...x_{n})=(...(x_{1}m+x_{2})m+x_{3})m+...)m+x_{n}
\end{equation}

All calculations are performed in the binary system and as a
result the $code (x_{1}...x_{n})$ is the binary notation of the
number $x_{1}x_{2}...x_{n}$. Let us estimate the time required for
calculation as in ( 1 ). Here and below the time will be measured
by the number of operations with single-bit words.

When calculating $(x_{1}m+x_{2})$ we obtain a number of length
$2\lceil\log m \rceil $ bits, and when calculating $((x_{1}m+x_{2})m+x_{3})$ ,
a number  $3\lceil\log m \rceil $ bits long and so on. When we calculate
these values we have at least to look through the words of length
of $2\lceil\log m \rceil,3\lceil\log m\rceil ,..., n \lceil \log m \rceil$.
So it takes not less than $c\: n^{2}\log m$
bit operations to calculate $code (x_{1}...x_{n})$ by (1). So one can see that time
per  letter is not less than $c\: n \log m$.

The main idea of our approach is very simple. First we propose a
new arrangement of brackets:
\begin{equation}
\begin{array}{ll}
code(x_{1}...x_{n})= (...((x_{1}m+x_{2})(m \cdot m)+(x_{3} m+x_{4}))((m \cdot m)(m \cdot m))+\\
+((x_{5}m+x_{6})(m\cdot m)+(x_{7}m+x_{8}))+...)
\end{array}
\end{equation}
When we use (2) most of the multiplications are carried out with short numbers.
So the total time of calculation is small.

Secondly, we propose to use a fast method of multiplication in (2).
We will use the Sch\"{o}nhage-Strassen method of multiplication which is the
fastest one (see [7, 8]). In this method the time $T(L)$ of multiplication of two
binary numbers with $L$ digits (and the time of division of a number with
$2L$ digits by a number with $L$ digits) is given by
\begin{equation}
T(L)=O(L \log L\log\log L), L\to \infty
\end{equation}

Let us estimate the time of calculations when  (2) is used. Calculation of
$(m\cdot m),(x_{1}m+x_{2}),(x_{3}m+x_{4}),...,(x_{n-1}m+x_{n})$
takes $(n/2)+1$ multiplications of numbers with $\lceil \log m
\rceil$ digits, calculation of $((m\cdot m)(m\cdot m)),(x_{1}
m+x_{2})(m\cdot m)+(x_{3}m+x_{4}),...$, $(x_{n-3}m+x_{n-2})(m\cdot
m)+(x_{n-1}m+x_{n})$ takes $(n/4)+1$ multiplications of numbers
with $2\lceil \log m \rceil$ digits, and so on. Using this and the
estimate (3) we can see that  the time of calculation of $code
(x_{1}...x_{n})$ by (2) is equal to
$$
\begin{array}{lll}
O((n/2)(\log m \log\log m \log\log\log m)+\\
(n/4)(2\log (2m) \log\log (2m) \log\log\log (2m)+...))=\\
O(n \enskip \log^{2} n \log\log n)
\end{array}
$$
So we can see that the time per letter is equal to $O(\log^{2} n
\log\log n)$.

Thus the proposed method takes $O(\log^{2} n \log\log n)$ bit operations
per letter instead of at least $n$ bit operations.

Note that our scheme is also valid  for the task of
calculation of values of any given polynomial. 

{\bf Claim 1}.Let $P(a)=y_{1}a^{n-1}+y_{2}a^{n-2}+... + y_{n}$ be
a polynomial and
$y_{1},y_{2},...y_{n}$ be integers, 
$m=\log (max \{|a|,|y_{1}|,...,|y_{n}|\}).$ The method of
calculation of the value $P(a)$ according to the formula
$$
\begin{array}{ll}
P(a)=((...((y_{1}a+y_{2})(a\cdot a)+(y_{3}a+y_{4}))((a\cdot a)(a\cdot a))+\\
(y_{5}a+y_{6})(a\cdot a)+(y_{7}\cdot a +y_{8}))...
\end{array}
$$
which uses the Sch\"{o}nhage-Strassen method of multiplication  takes not more than
$c\cdot n \cdot m \log^{2}(n\cdot m)\log\log (n\cdot m)$ bit operations when c is
constant, $n \to \infty$.

On the other hand, calculation by Horner scheme takes not less than
$const \cdot (n^{2} \cdot m)$ bit operations.

The proposed simple idea will be  used in this paper for fast ranking and
enumerative coding for the general case.
It is interesting that the method of "proper" arrangement of brackets is
a special case of divide-and-conquer principle (see the definition in [7]).

\section{Fast Ranking (or Encoding)}

Let $m \geq 2$ be an integer, $A=\{a_{1},a_{2},...,a_{m}\}$  the
alphabet and $A^{n}$  a set of words of length $n$ in the
alphabet $A$, where $n\geq 0$ is an integer. Every $S \subset
A^{n}$ is called a source. An enumerative code $\varphi$ is given
by two mappings $\varphi^{c}:S \to \{0,1\}^{t}$, where $t=\lceil
\log |S|\rceil$ and $\varphi^{d}: \varphi^{c}(S)\to S$ , so that
$\varphi^{d}(\varphi^{c}(s))=s$ for all $s \in S $ (here and
below, $|x|$ is the cardinality of $x$ if $x$ is a set, and the
length of $x$ if $x$ is a word). The map $\varphi^{c}$ is the
encoder and the map $\varphi^{d}$ is the decoder. For the sake of
simplicity  we identify every word with a certain number from the
interval [0,1]. For example, $0110=3/8$. Without loss of
generality it is assumed that the alphabet $A$ is a set of
integers from the interval  $[0,m-1]$, and we may apply the
lexicographic order to $A^{n}$.

Let us describe an enumerative code from [3]. Denote by
$N_{s}(x_{1}...x_{k})$ the number  of words which belong to $S$ and have the
prefix $x_{1}...x_{k}$, $k=1,2,...,n-1$. For $x_{1}x_{2}...x_{n} \in S$ define
\begin{equation}
code (x_{1}...x_{n}) = \sum_{i=1}^{n} \sum_{a<x_{i}}
N_{S}(x_{1}...x_{i-1} a)
\end{equation}

It is the code word for $x_{1}...x_{n}$. It should be noted that
there is a lot of interesting  cases where the formula (4) allows
to calculate the code using non-exponential memory size.

We give two examples of coding according to the formula (4). Both
are taken from [1-3].

The first example is the enumeration of
binary words with a given number of ones.  There is a source $S$
generating $n$-length binary words, $n>0$. There are $r$, $0\leq r
\leq n$ ones in each word $x$.

It's easy to see that
\begin {equation}
N_{s}(x_{1}...x_{k-1} 0)=
\left (\begin{array}{cc}
n-k \\
r - \sum_{i=1}^{k-1} x_{i}
\end{array} \right)
\end{equation}
Using  this formula and (4) we obtain

\begin{equation}
code (x_{1}...x_{n})= \sum_{k=1}^{n} x_{k}
\left (\begin {array}{cc}
n-k \\
r -\sum_{i=1}^{k-1} x_{i}
\end{array} \right)
\end{equation}

A time estimation of $ c \, n \log n \log\log n \, \:(c > 0)$ bit
operations per letter is obtained in [2] for the problem of
enumeration of binary words with a given number of ones.

In the second example the enumeration of permutations is used.
Let
$A$ be $\{1,2,...,n\}$. Given $x_{1}x_{2}...x_{n}$ and  $i$, $1
\leq i \leq n$, $r_{i}$ denotes the number of integers which, first,
are less than $x_{i}$, and, second, are situated to the right of
$i$. The relation (4) becomes
\begin{equation}
code (x_{1}...x_{n})=\sum_{i=1}^{n} r_{i}(n-i)!
\end{equation}

Using Horner's scheme we obtain
\begin{equation}
code(x_{1}...x_{n})=(...(r_{1}(n-1)+r_{2})(n-2)+r_{3})...)
\end{equation}
It is easy to estimate the time of calculation  by (8) which is not less than
$c \: n^{2}$ bit operations, where $c>0$ is constant. So the time per letter
equals $c\: n$.

In order to describe the proposed method we consider a source $S \subset A^{n},
n \geq 1$ and a word $x_{1}...x_{n} \in S$.

Let us define
\begin{equation}
\left.
\begin{array} {ll}
P(x_{1})= N(x_{1})/ |S|, P(x_{k}/x_{1}...x_{k-1})=
N(x_{1}...x_{k})/ N(x_{1}...x_{k-1}) \\
q(x_{1})= \sum_{a<x_{1}} P(a), q(x_{k}/x_{1}...x_{k})= \sum_{a<x_{k}}
P(a/x_{1}...x_{k-1}), k=2,...,n
\end{array} \right\}
\end{equation}
Clearly,
$$
\left.
\begin{array} {ll}
\sum_{i=1}^{n} \sum_{a<x_{i}} N(x_{1}...x_{i-a} a)= |S| (q(x_{1})+q(x_{2}/x_{1})P(x_{1})\\
+q(x_{3}/x_{1}x_{2}) P(x_{2}/x_{1})P(x_{1}) + q(x_{4}/x_{1}x_{2}x_{3})
P(x_{3}/x_{1}x_{2} P(x_{2}/x_{1}) P(x_{1}) +...)
\end{array} \right \}
$$ 
From this equality and (4) we obtain

\begin{equation}
code(x_{1}...x_{n})=|S|(q(x_{1})+q(x_{2}/x_{1})P(x_{1})+
q(x_{3}/x_{1}x_{2})P(x_{2}/x_{1})P(x_{1})+...)
\end{equation}

In short, the proposed method  may be described as follows: first,
use the proper arrangement of brackets in (10) and, second,  carry
out all calculations using rational numbers. For the sake of
simplicity we assume that $\log n$ is an integer. (In general
case we can add, for example, the letters $0$ to every word from
$S$ in order to make $\log n$ an integer. It does not affect $|S|$
and the complexity of the code.) The formal implementation of the
proper arrangement of brackets is:
\begin{equation}
\left.
\begin{array}{ll}
\rho_{1}^{0}=P(x_{1}), \rho_{2}^{0}=P(x_{2}/x_{1}),...,\rho_{n}^{0}=
P(x_{n}/x_{1}x_{2}...x_{n-1}) \\
\lambda_{1}^{0}=q(x_{1}), \lambda_{2}^{0}=q(x_{2}/x_{1}),...,
\lambda_{n}^{0}=q(x_{n}/x_{1}...x_{n-1})
\end{array} \right \}
\end{equation}

\begin{equation}
\left.
\begin{array}{ll}
\rho_{k}^{s}=\rho_{2k-1}^{S-1} \cdot \rho_{2k}^{S-1}, s=1,2,...,\log n;
k=1,2,...,n/2^{S}\\
\lambda_{k}^{S}= \lambda_{2k-1}^{S-1}+\rho_{2k}^{S-1} \cdot \lambda_{2k}^{S-1},
s=1,2,...,\log n; k=1,2,...,n/2^{S}
\end{array} \right \}
\end{equation}

All calculations are carried out using rational numbers and all
$\rho_{k}^{s}$ and $\lambda_{k}^{s}$ are fractions and presented as pairs of
integers. The Sh\"{o}nhage-Strassen method is used for multiplications.

As a result we have
$$
\begin{array}{ll}
\lambda_{1}^{\log n}=(q(x_{1})+q(x_{2}/x_{1})P(x_{1}))+(q(x_{3}/x_{1}x_{2})+\\
q(x_{4}/x_{1}...x_{3})P(x_{3}/x_{1}x_{2}))\cdot
(P(x_{1})P(x_{2}/x_{1}))+...
\end{array}
$$
We define the proposed code $\alpha^{c}$ as follows:

\begin{equation}
\alpha^{c}(x_{1}...x_{n})=|S| \cdot \lambda_{1}^{\log n}
\end{equation}

Now let us consider some examples.

First, we consider the ranking of binary words with a given number
of ones. Recall that
$$ \left (\begin {array}{cc}
t  \\  p
\end{array}  \right)  =
\left (\begin {array}{cc} t-1 \\
p-1
\end{array}
\right)  \cdot
\frac{t}{p}\enskip  ,\quad \left (
\begin  {array}{cc}  t\\
p
\end{array}
\right)=  \left  (\begin  {array}{cc}
t-1 \\
p
\end{array} \right) \cdot \frac{t}{t-p} $$
Let $\Delta$ be $0$ or
$1$. Combining the last equalities, we obtain

$$
\left (\begin {array}{cc}
t-1 \\
p- \Delta
\end{array} \right) /
\left (\begin {array}{cc}
t \\
p
\end{array} \right)  = \frac{\Delta \cdot p +(1-\Delta)(t-p)}{t}
$$
This equality and (9), (5) yield
\begin{equation}
P(x_{t}/x_{1}...x_{t-1})=\frac{x_{t}(k-\sum_{j=1}^{t-1}x_{j})+(1-x_{t})
(n-t+1-(k-\sum_{j=1}^{t-1}x_{j}))}{n-t+1}
\end{equation}

\begin{equation}
q(x_{t}/x_{1}...x_{t-1})=\frac{x_{t}(n-t+1-(k-\sum_{j=1}^{t-i} x_{j}))}
{n-t+1} ,
\end{equation}
$t=1,2,...,n.$

Let us give an example. Let $n=8, k=3$ and the word
$x_{1}x_{2}...x_{8}=01000101$. From (14), (15) and (11), (12)
we obtain
$$p(x_{1})=p(0)=\frac{0(3-0)+(1-0)(8-1+1-(3-0))}{8-1+1}=5/8$$
$$p(x_{2}/x_{1})=p(1/0)=\frac{1(3-0)+(1-1)(8-1+1-(3-0))}{8-2+1}=3/7$$
$$p(x_{3}/x_{1}x_{2})=p(0/01)=\frac{0(3-1)+(1-0)(8-3+1-(3-1))}{8-3+1}=4/6$$
$$p(x_{4}/x_{1}x_{2}x_{3})=\frac{0(3-1)+(1-0)(8-4+1-(3-1))}{8-4+1}=3/5$$
$$p(x_{5}/x_{1}x_{2}x_{3}x_{4})=p(0/0100)=\frac{0(3-1)+(1-0)(8-5+1-(3-1))}
{8-5+1}=2/4$$
$$p(x_{6}/x_{1}...x_{5})=p(1/01000)=\frac{1(3-1)+0(8-6+1-(3-1))}{8-6+1}=2/3$$
$$p(x_{7}/x_{1}...x_{6})=p(0/010001)=\frac{0(3-2)+(1-0)(8-7+1-(3-2))}{8-7+1}=1/2$$
$$p(x_{8}/x_{1}...x_{7})=p(1/0100010)=\frac{1(3-2)+(1-1)(8-8+1-(3-2))}
{8-8+1}=1/1$$
$$q(x_{1})=q(0)=0; q=(x_{2}/x_{1})=q(1/0)=\frac{1(8-2+1-(3-0))}{8-2+1}=4/7$$
$$q(x_{3}/x_{1}x_{2})=q(x_{4}/x_{1}...x_{3})=q(x_{5}...)=0$$
$$q(x_{6}/...)=q(1/01000)=\frac{1(8-6+1-(3-1))}{8-6+1}=1/3$$
$$q(x_{7}/...)=0, q(x_{8}/...)=q(1/0100010)=\frac{1(8-8+1-(3-2))}{8-8+1}=0$$

$$\rho_{1}^{0}=5/8, \rho_{2}^{0}=3/7, \rho_{3}^{0}=4/6, \rho_{4}^{0}=3/5,
\rho_{5}^{0}=2/4, \rho_{6}^{0}=2/3,\rho_{7}^{0}=1/2, \rho_{8}^{0}=1$$
$$\lambda_{1}^{0}=0,\lambda_{2}^{0}=4/7, \lambda_{3}^{0}=0,\lambda_{4}^{0}=0,
\lambda_{5}^{0}=0, \lambda_{6}^{0}=1/6,\lambda_{7}^{0}=0,\lambda_{8}^{0}=0$$
$$\rho_{1}^{1}=5/8 \cdot 3/7, \rho_{2}^{1}=4/6 \cdot3/5, \rho_{3}^{1}=2/4
\cdot 2/3, \rho_{4}^{1}=1/2 \cdot 1/2 \cdot 1$$
$$\lambda_{1}^{1}= 0+ 5/8 \cdot 4/7, \lambda_{2}^{1}=0+0, \lambda_{3}^{1}=
0+1/3 \cdot 2/4, \lambda_{3}^{1}=0+ \cdot 1/3 \cdot 2/4, \lambda_{4}^{1}=
0+0$$
$$\rho_{1}^{2}= 5/8 \cdot 3/7 \cdot 4/3 \cdot 3/5 ; \rho_{2}^{2}=2/4 \cdot
2/3 \cdot 1/2 \cdot 1/1$$
$$\lambda_{1}^{2}=5/8 \cdot 4/7 + 5/8 \cdot 3/7 \cdot 0= 5/8 \cdot 4/7$$
$$\lambda_{2}^{2}=1/3 \cdot 2/4 + 2/4 \cdot 2/3 \cdot 0= 1/3 \cdot 2/3$$
$$\lambda_{1}^{3}=5/8 \cdot 4/7 + 5/8 \cdot 3/7 \cdot 4/6 \cdot 3/5 \cdot 1/3
\cdot 2/4 = 20/56 + 1/56= 21/56$$

Observe that there are
$\left (\begin{array}{cc}
8\\
3 \end{array} \right )= 56$
binary words of the length $8$ with $3$ ones.
Thus, from (13) we obtain a code word:
$$
\alpha^{c}(01000101)=56 \cdot (21/56)=21$$
Of course, calculations according to the formula (6) give the  same result:
$code (01000101)=21$.
(For the sake of  clearness we carry out all calculations with decimal numbers
instead of binary ones).

Let us consider the enumeration of permutations. From the definition we obtain
$N_{s}(x_{1}...x_{k})=(n-k)!$; see also (7). The equalities (9), (11)-(13) yield

$$
\alpha^{c}(x_{1}x_{2}...x_{n})=$$
$$n!\Bigl (\Bigl (\frac{r_{1}}{n}+\frac{r_{2}}{n\cdot(n-1)}\Bigl )+\Bigl (\frac{1}{n} \cdot \frac{1}{n-1}\Bigl )
\Bigl (\frac{r_{3}}{n-2}+\frac{r_{4}}{(n-2)(n-3}\Bigl )+$$
$$+\Bigl (\Bigl (\Bigl (\frac{1}{n}\cdot \frac{1}{n-1} \Bigl )\cdot \Bigl (\frac{1}{n-2} \cdot\frac{1}{n-3}\Bigl )\Bigl )
\Bigl (\Bigl (\frac{r_{5}}{n-4}+\frac{r_{6}}{(n-2)(n-5)}\Bigl )+$$
$$+\Bigl (\frac{1}{(n-4)}\cdot \frac{1}{(n-5)}\Bigl )\Bigl (\frac{r_{7}}{n-6}+\frac{r_{8}}{(n-6)(n-7)}\Bigl )+...\Bigl )$$

In order to estimate the complexity of the method $\alpha^{c}$ we define
several values. Let, as before, $S\subset A^{n}$ be given. By definition $T$ is
the maximal time (in bit operations) for calculation of rational fractions
$N(x_{1}...x_{t+1})/N(x_{1}...x_{t})$, where $x_{1}...x_{n}\in S$,
$t=1,2,...,n-1$,  
$M$ is the size (in bits) of the program that is used
to compute
$$
\{ N(x_{1}...x_{t-1})/N(x_{1}...x_{t}); x_{1}...x_{n} \in S,t=1,2,...n-1\}
$$
and let $Q$ be the maximal denominator of  rational fractions
$$N(x_{1}...x_{t+1})/N(x_{1}...x_{t}), x_{1}...x_{n}\in S, t=1,...,n.$$
Let us define
\begin{equation}
\hat{Q}=max \{ |A|, Q\}
\end{equation}

{\bf Theorem 1}. {\it Let there be given an alphabet $A$, an integer $n$ and
$S\subset A^{n}$. The proposed method of encoding $\alpha^{c}$  has
the following properties:

i) $\alpha^{c}$ is correct, i.e. for every $x, y\in S \enskip \alpha^{c}(x) \neq
\alpha^{c}(y)$ and $\alpha^{c}(x)$ is an integer from $[0,|S|-1]$

ii) the time of encoding per letter is
$$ T+O(\log n \enskip \log \hat{Q} \log (n\enskip \log \hat{Q})\log\log (n \enskip \log \hat{Q}))$$
bit operations

iii) the memory size of the encoder is $M+O(n\enskip \log \hat{Q} \log n)$ bits. }

Proof. The claim i)  immediately follows from (10)-(13).

For the sake of  simplicity of the proof of ii) we assume that
$\log n$ and $\log \hat{Q}$ are integers. According to the
definition of $\hat{Q}$ and (14), (15) we can see that the
notation of every $P(\enskip )$ and $q(\enskip )$ uses $2 \log
\hat{Q} $ bits ($\log\hat{Q}$ bits for the numerator and
$\log\hat{Q}$ bits for the denominator). That is why the
calculation of $\rho_{k}^{1}, k=1,2,...,n/2$ according to (12)
takes $2(n/2)$ multiplications of numbers of the length
$\log\hat{Q}$ bits and the calculation of $\lambda_{k}^{1},
k=1,...,n/2$ according to (12) and the   formula $a/b + c/d =(ad +
bc)/(bd)$ takes $3(n/2)$ multiplications of numbers with the
length $\log\hat{Q}$ bits. The calculations of
$\rho_{k}^{2},\lambda_{k}^{2}, k=1,2,...,n/4$ take $5(n/4)$
multiplications of numbers of the length $2\log\hat{Q}$ bits each.
Similarly, the calculation of $\rho_{k}^{i},\lambda_{k}^{i},
k=1,2,...,n/2^{i}$ takes $5(n/2^{i})$ multiplications of numbers
with the length $2^{i}\log n$ bits. From (3) we obtain that the
general time of calculations is:
$$(5 n/2)O(\log\hat{Q}\enskip\log\log\hat{Q}\enskip\log\log\log\hat{Q}) +$$
$$(5 n/4)O(2\enskip\log\hat{Q}\enskip\log (2\log\hat{Q}) \log\log (2\enskip\log\hat{Q})+ ...$$
$$(5 n/2^{i})O(2^{i}\enskip\log\hat{Q}\enskip \log(2^{i}\log\hat{Q})\log\log(2^{i}
\log\hat{Q})+$$
$$...+ 5\cdot O(n\enskip\log\hat{Q}\enskip\log (n \enskip\log\hat{Q})\log
\log (n \enskip\log\hat{Q})$$
It is easy to see that the last value is not more than
$$
O(n \enskip\log n \enskip\log\hat{Q})\enskip\log (n \enskip\log\hat{Q}) \log\log (n \enskip\log\hat{Q}))$$
bit operations. It yields
\begin{equation}
O(\log n\enskip\log\hat{Q}\enskip\log
(n\enskip\log\hat{Q})\log\log (n \enskip\log\hat{Q}))
\end{equation}
bit operations per letter  for calculation of $\lambda_{1}^{\log
n}$. In order to obtain $\lambda^{c}(x_{1}...x_{n})$ we should
calculate the product $|S|\lambda_{1}^{\log n}$, see (13). $S$ is
a subset of $A^{n}$, so $|S|\leq |A|^{n}$ and a binary notation of
the numbers $|S|$ and $\lambda_{1}^{\log n}$ takes not more than
$n\cdot\log |A|$ bits. From (3) we can see that  the time of
calculation of $|S|\lambda_{1}^{n}$ is equal to $O(n\enskip \log
|A|\log (n\enskip |A|)\log\log (n \enskip |A|)$ bit operations per
letter. From this, (17), and (16) we obtain ii).

In order to estimate the size of the encoder program, note that when it
calculates $\lambda_{k}^{i}, \rho_{k}^{i}$ it can store only $\lambda_{k}^{i-1},
\rho_{k}^{i-1}, i=2,...,\log n$, the same memory is used to store $\{\lambda_{k}^{i-1},
\rho_{k}^{i-1}; k=1,...,n/2^{i-1}\}$ and $\{\lambda_{k}^{i}, \rho_{k}^{i};
k=1,...,n/2^{i}\}$. From this and the definitions of $m$ and $\hat{Q}$ we
can easily obtain iii). Theorem 1 is proved.

\section{Fast Decoding}

First, we describe the general scheme of decoding not taking into account the time
of calculation. Let an alphabet $A=\{0,1,...,m-1\}$ and a source $S\subset A^{n}$
be given and let $\hat{x}=x_{1}x_{2}...x_{n}$ be a word from $S$ and
$y=\alpha^{c}(\hat{x})$ be the encoded word  $\hat{x}$.

In order to decode $\hat{y}$ we consider $y_{1}=y/|S|$
as a rational number and first find $i_{1}$ satisfying the
inequalities

\begin{equation}
\lambda_{i_{1}}^{0}\leq \hat{y}_{1} < \lambda_{i_{1}+1}^{0}
\end{equation}

From these inequalities it follows that the first letter of the
encoded word is $i_{1}$: $x_{1}=i_{1}$. After that we calculate
$$
\hat{z}=(y_{1}- \lambda_{i_{1}}^{0})/\rho_{i_{1}}^{0}
$$
where $\hat{z}$ is a rational number, and find $i_{2}$ complying with the
inequalities

\begin{equation}
\lambda_{i_{2}}^{0}\leq\hat{z} < \lambda_{i_{2}+1}^{0}
\end{equation}
If follows that the second letter is $i_{2}$.

Of course, we could use this way  to find the third letter, then
the fourth one, etc. But we use a more complicated way which  will
give a possibility to operate with short numbers. We calculate
$\lambda_{1}^{1}$ according to (12). (It is possible because now
$x_{1}$ and $x_{2}$ are known now.) After that we calculate

\begin{equation}
y_{2}=(y_{1}-\lambda_{1}^{1})/\rho_{1}^{1}
\end{equation}
and find letters $x_{3},x_{4}$ using $y_{2}$ as we have found $x_{1},
x_{2}$ using $y_{1}$. Then we calculate $\lambda_{2}^{1}$ using $x_{3}$
and $x_{4}$ and $\lambda_{1}^{2}$ using $\lambda_{1}^{1}, \lambda_{2}^{1}$
and $\rho_{1}^{1}$, see (12). And so on.

The point is that when we carry out calculations (18)-(20) we can use
only estimations of $y_{1}, y_{2},\lambda_{i}^{1}, \rho_{i}^{1}$,
etc, which are based on the few leading digits. More exactly, we
will use two estimations for every value which are an upper bound
and a lower one.

In order to give the exact definition, first we define several
auxiliary values. Let $p/q$ be a rational number
represented as a pair of the integers $p,q$, $0 < p \leq q$, and
let $t \geq 1$ be an integer. We define two functions
$\varphi_{t}^{+}(p/q)$ and $\varphi_{t}^{-}(p/q)$ as follows. Let
$l= \lfloor \log q \rfloor$, and $(q_{l}q_{l-1}...q_{0})$ and
$(p_{l}...p_{0})$ be  binary representations of $q$ and $p$,
correspondingly. Then
$$
\varphi_{t}^{+}(p/q)=\Bigl(\sum_{i=l-t+1}^{l}p_{i}2^{i}+ 2^{l-t}\Bigr)/
\Bigl(\sum_{i=l-t+1}^{l} q_{i} 2^{i}\Bigr)$$
$$
\varphi_{t}^{-}(p/q)=\Bigl(\sum_{i=l-t+1}^{l} p_{i}2^{i}\Bigr)/
\Bigl(\sum_{i=l-t+1}^{l} q_{i}2^{i}+2^{l-t}\Bigr)
$$
For example, $\varphi_{3}^{+}(5/17)=3/8$, $\varphi_{3}^{-}(5/17)=2/9$.

We will need the following simple bounds.

{\bf Lemma.} Let $p,q,t$ be integers , $0<p\leq q$, $t>2$. Then
\begin{equation}
0 \leq \varphi_{t}^{+}(p/q)-p/q < 2^{2-t}
\end{equation}
\begin{equation}
0\leq p/q -\varphi_{t}^{-}(p/q)< 2^{2-t}
\end{equation}

Proof. It's easy to see that if $x< 1/2$ then
\begin{equation}
\frac{1}{1-x} <1+2x,\enskip \frac{1}{1+x} > 1-x
\end{equation}

Theae bounds immediately follow from well known equalities
$$(1-x)^{-1}=1+x+x^{2}+...=1+x+x^{2}/(1-x) =1+x(1+x/(1-x))$$
$$(1+x)^{-1}=1-x+x^{2}-...=1-x+x^{2}(1-x+x^{2}-...)$$

The following sequence of inequalities gives the bound (21):
$$\varphi_{t}^{+}(p/q)\leq \frac{p+2^{l-t}}{q-2^{l-t}} <p/q
(1+2^{l-t}/p)(1+2\cdot 2^{l-t}/q)=$$
$$p/q+2^{l-t}/q+2\cdot 2^{l-t}/q+2^{l-t}/q \cdot 2\cdot^{l-t}/q <p/q+ 2^{-t}+$$
$$2\cdot 2^{-t}+ 2^{-2t} < p/q+ 4\cdot 2^{-t}$$
Here we use (23) and the obvious inequality $2^{l}\leq q$.


Let us proceed with the description of the method of decoding.
Let, as before, an alphabet $A=\{ 0,1,...,m-1\}$ and a source
$S\subset A^{n}$ be given.

As before, let $Q$ be the maximal denominator of the rational
numbers $N(x_{1}...x_{t+1})/N(x_{1}...x_{t})$, $x_{1}...x_{t+1}\in
S$, $t=1,2,...,n-1.$ From this definition and (11), (12) we
immediately obtain that
 the denominators
of the rational fractions $\rho_{i}^{s}$ and $\lambda_{i}^{s}$ not exceeding
$Q^{s}$, for all $s=1,...,\nu$; $k=1,...,n/2^{s}$. Let
\begin{equation}
h=\lceil \log Q \rceil +3
\end{equation}

We will give the definition by induction on $n$. First, let
$n=2$. For every value $\lambda_{j}^{i}$ we define the upper and the lower
estimations, $\lambda^{+}(i,j)$ and $\lambda^{-}(i,j)$.

Let the decoder calculate
\begin{equation}
\lambda^{+}(1,1)=\varphi_{2h}^{+}(y/|S|),\enskip \lambda^{-}(1,1)=
\varphi_{2h}^{-}(y/|S|)
\end{equation}
\begin{equation}
\lambda^{+}(0,1)=\varphi_{h}^{+}(\lambda^{+}(1,1)), \enskip \lambda^{-}(0,1)=
\varphi_{h}^{-}((\lambda^{-}(1,1))
\end{equation}

Then it  finds $i_{1}$ complying with the inequalities
\begin{equation}
q(i_{1}) \leq \lambda^{+}(0,1), q(i_{1}+1) > \lambda^{-}(0,1)
\end{equation}

We use these inequalities instead of (23). But here the decoder carries out
calculations with $(h+3)$ - length words instead of the whole binary notations
of $y$ and $|S|$. The inequalities (27) mean that the first letter of coded words is
$i_{1}$: $x_{1}=i_{1}$.Let us define
\begin{equation}
\begin{array}{ll}
\lambda^{+}(0,2)=\varphi_{h}^{+}((\lambda^{+}(1,1)-\lambda_{1}^{0})/
\rho_{1}^{0})\\
\lambda^{-}(0,2)=\varphi_{h}^{-}((\lambda^{-}(1,1)-\lambda_{1}^{0})/
\rho_{1}^{0})
\end{array}
\end{equation}
and find $i_{2}$ complying with inequalities
$$
q(i_{2})\leq \lambda^{+}(0,2), \enskip q(i_{2}+1) > \lambda^{-}(0,2)$$
It means that the second letter is $i_{2}$: $x_{2}=i_{2}$.

Let now $n$ be greater than $2$. In order to use the divide-and-conquer principle
we define
\begin{equation}
\begin{array}{cc}
\lambda^{+}(\lceil\log n\rceil, 1)=\varphi_{nh}^{+}(y/|S|)\\
\lambda^{-}(\lceil\log n\rceil,1)=\varphi_{nh}^{-}(y/|S|)\\
\lambda^{+}(\lceil\log n\rceil -1,1)=\varphi_{\lceil nh/2\rceil}^{+}(\lambda_{-}^{+}(\lceil\log n\rceil,1))\\
\lambda^{-}(\lceil\log n\rceil- 1,1)=\varphi_{\lceil nh/2\rceil}^{-}(\lambda(\lceil\log n\rceil, 1))
\end{array}
\end{equation}

Then the decoder finds $x_{1},x_{2},...,x_{\lceil n/2\rceil}$ using
$\lambda^{+}(\lceil \log n \rceil -1,1)$, $\lambda^{-}(\lceil \log n\rceil -1,1)$
and calculates $\lambda_{1}^{\lceil \log n \rceil -1}$ (see (11), (12)). After that
the decoder calculates
\begin{equation}
\begin{array}{ll}
\lambda^{+}(\lceil\log n\rceil -1,2)=\varphi_{\lceil n h/2\rceil}^{+}
((\lambda^{+}(\lceil \log n \rceil, 1)- \lambda_{1}^{\lceil\log n\rceil -1})/
\rho_{1}^{\lceil\log n\rceil - 1})\\
\lambda^{-}(\lceil\log n\rceil -1,2)=\varphi_{\lceil n h/2\rceil}^{-}
((\lambda^{-} (\lceil \log n \rceil, 1)- \lambda_{1}^{\lceil\log n\rceil -1})/
\rho_{1}^{\lceil\log n\rceil - 1})
\end{array}
\end{equation}
and uses this pair in order to find  $x_{\lceil n/2\rceil+1},
x_{\lceil n/2\rceil+2},...,x_{n}$. So (29) and (30) give  a
possibility to decode the   $n$-letter word as two words of length
$\lceil n/2\rceil$ and $(n-\lceil n/2\rceil)$, correspondingly.

In order to give an example of the decoding, let us consider the previous example.
Let, as before, $S$ be a set of all binary words of length $8$ and each of them
has $3$ ones. Let the proposed method be applied for decoding
of the word $010101=(21)_{10}$.

According to the description, first, the decoder finds
$$
\lambda^{+}(\lceil\log n \rceil,1)=\lambda^{+}(3,1)=\varphi_{48}^{+}(25/56)=
(21\cdot 2^{42}+1/56 \cdot 2^{42})$$
$$
\lambda^{-}(\lceil\log n \rceil, 1)=\lambda^{-}(3,1)=\varphi_{48}^{-}(25/56)=
(21\cdot 2^{42}/56 \cdot 2^{42} +1)$$
and
$$
\lambda^{+}(2,1)=\varphi_{21}^{+}(21\cdot 2^{42}+1/56\cdot 2^{42})=
(21 \cdot 2^{23} +1/56\cdot 2^{23})$$
$$\lambda^{-}(2,1)=(21 \cdot 2^{23}/56 \cdot 2^{23} +1)$$

Using $\lambda^{+}(2,1)$ and $\lambda^{-}(2,1)$ the decoder should find
$x_{1},x_{2},x_{3},x_{4}$. According to the algorithm, it calculates
$$
\begin{array}{ll}
\lambda^{+}(1,1)=\varphi_{12}^{+}(21 \cdot 2^{23} +1/56 \cdot 2^{23})=
(21 \cdot 2^{7}+1/56 \cdot 2^{7})\\
\lambda^{-}(1,1)=(21 \cdot 2^{7} /56 \cdot 2^{7}+1)
\end{array}
$$

This pair encodes $x_{1},x_{2}$. After that the decoder finds
$$\lambda^{+}(0,1)=\varphi_{6}^{+}(21\cdot 2^{7}+ 1/56\cdot 2^{7})=22/56$$
$$\lambda^{-}(0,1)=\varphi_{6}^{-}(21\cdot 2^{7}+1/56\cdot 2^{7})=21/57$$
which  encode $x_{1}$. For given $S$    $q(0)=0, q(1)=5/8$ (see the
example of coding), $x_{1}=0$ because
$$
0=q(i_{1}) \leq \lambda^{+}(0,1)=22/56$$
$$
21/57=\lambda^{-}(0,1) < q(i_{1}+1)=1$$
(see (22)). According to (24) the decoder calculates
$$
\lambda^{+}(0,2)=\varphi_{6}^{+}((21\cdot 2^{6}+1/56\cdot 2^{7} -0)/(5/8)
=169/280$$
$$
\lambda^{-}(0,2)=168/281$$
Thus, $x_{2}=1$ because $q(0/0)=0, q(1/0)=4/7$ and, obviously,
$$ 4/7 \leq 169/280=\lambda^{+}(0,2), \lambda^{-}(0,2)= 168/281 <1.$$
After finding $x_{1}=0$ and $x_{2}=1$ the decoder calculates $\lambda_{1}^{1}=
5/14, \rho_{1}^{1}=15/56$  and finds $\lambda^{+}(1,2), \lambda^{-}(1,2)$
according to (24). It gives $x_{3}=0, x_{4}=0$ and so on.

The next theorem characterizes properties of the proposed method of decoding
which we denote as $\alpha^{d}$.

{\bf Theorem 2.} {\it Let there be an alphabet $A$, an integer $n$ and
$S\subset A^{n}$. Then the proposed method of decoding $\alpha^{d}$ has  the
following properties:

i) $\alpha^{d}$ is correct, i.e. for every $x \in S$
$$\alpha^{d}(\alpha^{c}(x))=x$$

ii) the time of decoding per letter is
$$ T+O(\log \hat{Q}(\log n\enskip \log (n\enskip \hat{Q}))\log\log (n \enskip\hat{Q})$$

iii) the memory size of the decoder is
$$M + O(n \enskip \log \hat{Q} \log n)$$

(here $T,M$ and $\hat{Q}$ are defined as in Theorem 1.) }

Proof. First, we estimate the speed of decoding. As it follows from the algorithm
every operation of multiplication for calculation of $\lambda_{j}^{i}$
corresponds to two divisions when the decoder calculates $\lambda^{+}(i,j)$ and
$\lambda^{-}(i,j)$ according to (28)-(30). The time of divisions in (28)-(30)
is proportional to the time of multiplications, see (12). So the time
of calculations of
$\lambda^{+}(i,j)$ and $\lambda^{-}(i,j)$ is equal to the time of calculations of
$\lambda_{j}^{i}$ within a multiplicative constant. It is easy to see that the
time of finding $i_{1}=x_{1}$, $i_{2}=x_{2}...$, according to (27) does not change
the asymptotical estimation of the time of decoding.

Let us estimate the memory size. For this purpose we note that the
decoder can use  the same memory size for decoding the first
letters $x_{1}...x_{\lceil n/2 \rceil }$ and the letters
$x_{\lceil n/2\rceil +1},...,x_{n}$. From this fact it immediately
follows that  the decoder can use such a memory size as does the
encoder and it gives the same estimation for the memory size.

Now we will show that $\alpha^{d}$ is a correct method of decoding. For this purpose
for a letter $x_{j}$, $j=1,...n$, we estimate the values
$$
|\lambda^{+}(0,j)- q(x_{j})|\enskip \mbox{and}\enskip |q(x_{j})-\lambda^{-}(0,j)|
$$
It is important because the decoder decides that letter $x_{j}$ sould be
decoded as $i_{j}$ if the inequalities
\begin{equation}
q(i_{j}) \leq \lambda^{+}(0,j), \enskip q(i_{j}+1) \geq \lambda^{-}(0,j)
\end{equation} are valid. As it follows from (11), the (31) 
are equal to
\begin{equation}
\lambda_{i_{j}}^{0} \leq \lambda^{+}(0,j),\enskip\lambda_{i_{j}+1}^{0}\geq\lambda^{-}(0,j)
\end{equation}
and these inequalities should be valid for one $i_{j}$ in the case of $x_{j}=i_{j}$.
And this property should be valid for all letters $x_{1},x_{2},...,x_{n}$ and for all
$x_{1}...x_{n} \in S$.

By definition $Q$ is the maximal denominator of the rational fractions
$$
P(x_{t+1}/x_{1}...x_{t})=N(x_{1}...x_{t+1})/N(x_{1}...x_{t})
$$
$t=1,...,n-1$; $x_{1}...x_{n} \in S$. It means that for every $x_{1}...x_{t}$,
$i,j \in A$; $i\neq j$.
\begin{equation}
\begin{array}{cc}
|q(i/x_{1}...x_{t})-q(j/x_{1}...x_{t})| \geq 1/Q\\
q(i/x_{1}...x_{t}) \geq 1/Q \,;
\end{array}
\end{equation}
see (9). From the definition (11) we obtain
$$
|\lambda_{i}^{0} -\lambda_{j}^{0}| \geq 1/Q
$$

From  this inequality and (32) we can see that the inequalities
$$
\begin{array}{cc}
0 \leq \lambda^{+}(0,j) -\lambda_{j}^{0} < 1/Q \\
0 \leq \lambda_{j}^{0} -\lambda^{-}(0,j) < 1/Q
\end{array}
$$
guarantee the correctness of decoding. We will prove only the
first pair of inequalities because the second one can be proved in
the same way. We will investigate the value $\lambda^{+}(0,n)-
\lambda_{n}^{0}$ because it can be easily seen from a proof that
the possible error is maximal for the last letter $x_{n}$. First,
we notice that the inequality
$$ \lambda^{+}(0,j)-\lambda_{j}^{0} \geq 0$$
is immediately obtained from the definition of $\varphi^{+}(\enskip )$ and
$\lambda^{+}(\enskip )$, see (25)-(30). Now we have to prove that
\begin{equation}
\lambda^{+}(0,n)-\lambda_{n}^{0} < 1/Q
\end{equation}
We define
\begin{equation}
\varepsilon (i,j)=\lambda^{+}(i,j)-(\lambda^{+}(i+1,j/2)-\lambda_{j-1}^{i})/
\rho_{j-1}^{i})
\end{equation}
$i=0,...,j=n/2^{i}$. In fact, $\varepsilon (i,j)$ is an error arising from using
$\varphi_{2^{ih}}^{+}((\lambda^{+}(i+1,j/2)-\lambda_{j-1}^{i})\rho_{j-1}^{i})$
instead of the value $(\lambda^{+}(i+1,j/2)-\lambda_{j-1}^{i})/\rho_{j-1}^{i}$.
The following train of expressions is valid:
$$
\lambda^{+}(0,n)=\varphi_{h}^{+}((\lambda^{+}(1,n/2)-\lambda_{n-1}^{0})/ \rho_{n-1}^{0}=
\varepsilon (0,n)+(\lambda^{+}(1,n/2)-\lambda_{n-1}^{0})/ \rho_{n-1}^{0} =$$
$$\varepsilon (0,n)+
(\varphi_{2h}^{+}((\lambda^{+}(2,n/4)-\lambda_{n/2-1}^{1})/ \rho_{n/2-1}^{1})-
\lambda_{n-1}^{0})/ \rho_{n-1}^{0}=$$
$$\varepsilon (0,n)+((\varepsilon (1,n/2)+(\lambda^{+}(2,n/4)-\lambda_{n/2}^{1})/
\rho_{n/2-1}^{1})-\lambda_{n-1}^{0})/\rho_{n-1}^{0}=$$
$$=\varepsilon (0,n)+ \varepsilon (1,n/2)/ \rho_{n-1}^{0}+((\varphi_{4h}^{+}
(\lambda^{+} (3,n/8)-\lambda_{n/4-1}^{2})/ \rho_{n/4-1}^{2}
-\lambda_{n/2}^{1})/$$
$$\rho_{n/2-1}^{1})-\lambda_{n-1}^{0})\rho_{n-1}^{0}=
\varepsilon (0,n)+ \varepsilon (1,n/2)/ \rho_{n-1}^{0}
+ \varepsilon (2,n/4)/
(\rho_{n-1}^{0} \cdot \rho_{n/2}^{1})+$$
$$(...(\varphi_{8h}^{+}(\lambda^{+}(4,n/2^{4})-\lambda_{n/8-1}^{3})/
\rho_{n/8-1}^{3} -\lambda_{n/4-1}^{2})/\rho_{n/4-1}^{2}
-\lambda_{n/2}^{1})/\rho_{n/2-1}^{1})-\lambda_{n-1}^{0})/\rho_{n-1}^{0}
 $$  $$ = ...
=\varepsilon (0,n)+ \varepsilon (1,n/2)/\rho_{n-1}^{0}+
\varepsilon (2,n/4)/ (\rho_{n-1}^{0} \cdot \rho_{n/2-1}^{1})+
\varepsilon (3,n/8)/$$
$$(\rho_{n-1}^{0} \cdot \rho_{n/2-1}^{1} \cdot
\rho_{n/4-1}^{2})+...
+\varepsilon (\log n,1)/(\rho_{n-1}^{0} \cdot \rho_{n/2-1}^{1}...
\rho_{1}^{\log n-1}) +$$
$$+((...(\lambda_{1}^{\log n} -\lambda_{1}^{\log n-1})/\rho_{1}^{\log n-1} -
\lambda_{3}^{\log -2})/ \rho_{3}^{\log n-2} -
\lambda_{7}^{\log n-3})/ \rho_{7}^{\log n-3} - ...)/\rho_{n-1}^{0}
$$
So we obtain
$$\lambda^{+}(0,n)= \sum_{i=0}^{\log n} \varepsilon (i,n \cdot 2^{-i})/
\prod _{j=0}^{i=1} \rho_{n 2^{-j}-1}^{j} +$$
$$+(...(\lambda_{1}^{\log n} -\lambda_{1}^{\log n-1})/ \rho_{1}^{\log n-1} -
\lambda_{3}^{\log n-2})/\rho_{3}^{\log n-2}-...)/\rho_{n-1}^{0}$$

Using the definition (12) of $\lambda_{j}^{i}$ we obtain
$$\lambda^{+}(0,n)= \sum_{i=0}^{\log n} \varepsilon (i,n 2^{-i})/
\prod _{i=0}^{i-1} \rho_{n2^{-j}-1}^{j} +\lambda_{n}^{0}$$
Thus,
$$
\lambda^{+}(0,n)-\lambda_{n}^{0}= \sum_{i=0}^{\log n} \varepsilon (i,n 2^{-i})/
\prod _{i=0}^{i-1} \rho_{n2^{-j}-1}^{j}
$$
This equality and (33), (12) yield
$$
\lambda^{+}(0,n)-\lambda_{n}^{0} \leq \varepsilon
(0,n)+\varepsilon (1,n/2)Q+ \varepsilon (2,n/4)Q^{3} +\varepsilon
(3,n/8)Q^{7}+ $$ $$...+ \varepsilon (\log n,1)Q^{2^{n}-1}$$ The
claim of the Lemma, (28), (29) and (35) yield
$$\varepsilon (i,j) < 2^{2-2^{i}h}$$
The last two inequalities and (24) give us
$$\lambda^{+}(0,n)-\lambda_{n}^{0}<4(1/8 Q+Q/(8Q)^{2}+Q^{3}/(8Q)^{4}+...)$$
Hence, we obtain the inequality
$$
\lambda^{+}(0,n)-\lambda_{n}^{0} <1 /Q
$$
which completes the proof of (34). Theorem 2 is proved.

\newpage
\begin{center}
{\Large \bf References}
\end{center}

\begin{enumerate}

\item{\it Reingold E.M., Nievergelt J., Deo N.}, "Combinatorial Algorithms.
Theory and Practice". Prentice-Hall, Inc., 1977.

\item{\it Krichevsky R.}, "Universal Compression and Retrieval",
Kluwer Academic Publishers, 1994.

\item{\it Cover T.M.} "Enumerative Source Encoding". IEEE Trans. Inform Theory,
vol. IT-19, pp. 73-77, Juan. 1973.

\item{\it Lynch T.Y.} Sequence time coding for data compression.// Proc.
IEEE, v.54, pp.1490-1491, 1966.

\item{\it Davisson L.D.} Comments on "Sequence time coding for data compression".//
Proc. IEEE, v.54, p.2010, 1966.

\item{\it Babkin V.F.} "A method of universal coding with non-exponent labour
consumption" Probl. Inform. Transmission, v.7, pp. 13-21, 1971.

\item{\it Aho A.V., Hopcroft L.E., Ullman J.D.} "The Design and Analysis
of Computer  Algorithms".  Addison-Wesley.  Publishing
Company,1976.

\item{\it  Knuth D.E.}   "The art of computer programming." Vol.2.
Addison Wesley, 1981.
\
\end{enumerate}

\end{document}